\documentclass[pra]{revtex4-1}

\usepackage{amsmath,amssymb,braket}
\usepackage{graphicx}

\usepackage{sistyle}
\usepackage[utf8]{inputenc}

\begin{document}

\title{A Hidden Markov Model of atomic quantum jump dynamics in an optically probed cavity}

\author{S. Gammelmark}
\affiliation{Department of Physics and Astronomy, University of Aarhus, Ny
  Munkegade 120, DK-8000 Aarhus C, Denmark.}

 \author{W. Alt}
 \affiliation{Institut f\"ur Angewandte Physik der Universit\"at Bonn, Wegelerstrasse 8, 53115 Bonn, Germany}

 \author{T. Kampschulte}
 \affiliation{Institut f\"ur Angewandte Physik der Universit\"at Bonn, Wegelerstrasse 8, 53115 Bonn, Germany}

 \author{D. Meschede}
 \affiliation{Institut f\"ur Angewandte Physik der Universit\"at Bonn, Wegelerstrasse 8, 53115 Bonn, Germany}

  \author{K. M{\o}lmer}
\affiliation{Department of Physics and Astronomy, University of Aarhus, Ny
  Munkegade 120, DK-8000 Aarhus C, Denmark.}

\date{\today}
\begin{abstract}
We analyze the quantum jumps of an atom interacting with a cavity field. The strong atom-field interaction makes the cavity transmission depend on the time dependent atomic state, and we present a Hidden Markov Model description of the atomic state dynamics which is conditioned in a Bayesian manner on the detected signal. We suggest that small variations in the observed signal may be due to spatial motion of the atom within the cavity, and we represent the atomic system by a number of hidden states to account for both the small variations and the internal state jump dynamics. In our theory, the atomic state is determined in a Bayesian manner from the measurement data, and we present an iterative protocol, which determines both the atomic state and the model parameters. As a new element in the treatment of observed quantum systems, we employ a Bayesian approach that conditions the atomic state at time $t$  on the data  acquired both before and after $t$ and we show that the state assignment by this approach is more decisive than the usual conditional quantum states, based on only earlier measurement data.
\end{abstract}


\maketitle

\section{Introduction}

\noindent
Cavity QED experiments offer wide possibilities to vary interaction strengths, resonance conditions and dissipation rates, and thus provide ideal demonstrations of the dynamical evolution of single quantum systems \cite{Haroche-book}. The combined field and atom degrees of freedom allow, on the one hand, probing of a cavity field by transmission and detection of the state of atoms passing through the cavity \cite{ENS-exp}, and, on the other hand, probing of the quantum state of atoms located inside a cavity by detection of the transmitted field \cite{reick2010}. Recent experiments with superconducting circuit elements and resonators have shown similar measurement capabilities to  monitor over time the quantum state of individual quantum systems \cite{Murch-13,Groen-13,Sun-13}.

We have recently \cite{reick2010} shown how the incoherent jumping of one or two atoms between different hyperfine ground states can be monitored due to the different cavity transmission levels associated with the atomic states. Rather than inferring the atomic states directly from the noisy value of the transmission signal, a Bayesian approach was used, where the measured signal continuously updates our probabilistic knowledge based on the previous measurements. This was shown to provide more decisive predictions about the state of the system.  An extension of the Bayesian method to incorporate different unknown physical parameters of the light-atom interacting system has also been demonstrated \cite{Brakhane-12}.

In this paper we extend the modelling of the system to condition our estimation of the state of the atom at a given time $t$, not only on the optical measurement data acquired up to time $t$ but also by the data acquired later. This is a well developed procedure, known as smoothing, in the analysis of Hidden Markov Models (HMM), where methods also exist to combine the state estimation with the estimation of the physical model parameters from the available data.

The paper is organized as follows. In Sec. II we give a  description of the experimental set-up and the data obtained. In Sec III, we present a summary of the general Bayesian method applied in the paper and of the formalism needed to provide the atomic state based on both past and future measurements. In Sec. IV, we discuss how system parameters such as atomic transition probabilities and optical signal distributions can be efficiently determined in the HMM framework. In Sec. V, we apply our formalism to experimental data, and we show that the atomic state and model parameters are indeed more decisively determined by the full time analysis. In Sec. VI, we conclude with a discussion of the results and some further perspectives.

\section{Description of experiments}

\begin{figure}[!h]
  \centering
  \includegraphics[width=0.3\textwidth]{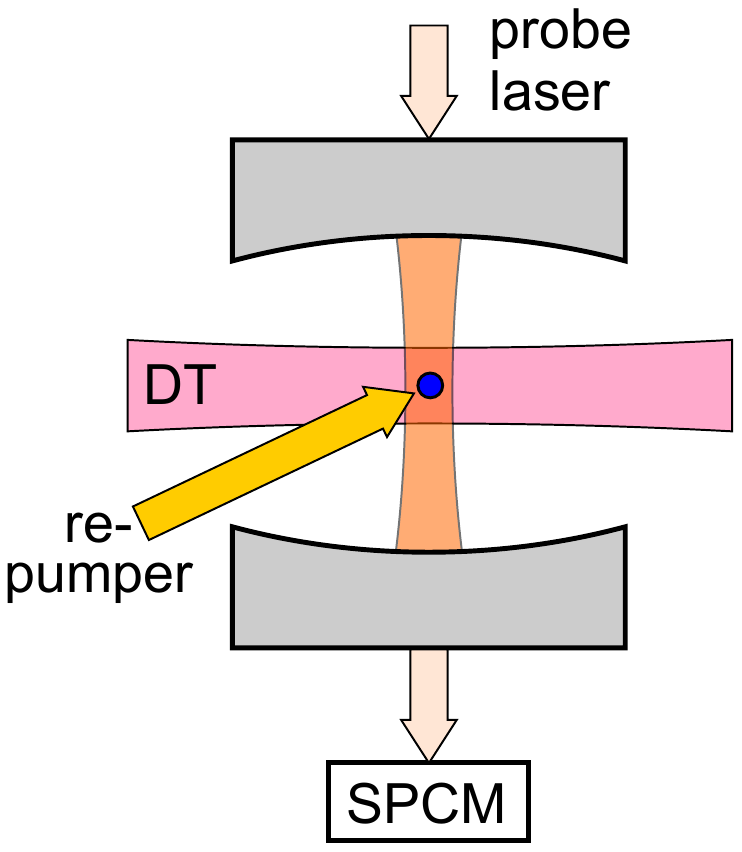}
  \caption{A single Caesium atom is held by a dipole trap (DT) inside the mode of a high-finesse optical cavity. The transmission of a probe laser is observed with a single-photon counting module (SPCM).}
  \label{fig:setup}
\end{figure}

In the experiment, we couple single neutral Caesium atoms to the mode of an optical high-finesse cavity, see Fig.~\ref{fig:setup}. Very few Caesium atoms are collected and laser-cooled in a high-gradient magneto-optical trap, and subsequently transferred into a far-detuned standing-wave optical dipole trap. By analyzing a fluorescence image we make sure that only a single atom is loaded~\cite{karski2009}. Using the dipole trap as an optical conveyor belt~\cite{kuhr2001}, we transport the atom to the center of the mode of an optical cavity with a very high finesse of $10^6$~\cite{khudav2008}. The small mode volume of the cavity (length 160\,$\mu$m, mode waist radius 23\,$\mu$m) leads to an effective atom-cavity coupling rate of $g\approx 2\pi\cdot 9$\,MHz~\cite{reick2010}, which is larger than the atomic dipole decay rate $\gamma = 2\pi\cdot2.6$\,MHz and the cavity field decay rate $\kappa = 2\pi\cdot0.4$\,MHz. These parameters place the system in the strong-coupling regime, where the single-atom cooperativity $g^2/(2\kappa\gamma)\approx 40$ is larger than one, and the absorptive and dispersive effects of a single atom lead to a nomal mode splitting of the cavity resonance (vacuum-Rabi-Splitting) and can thus strongly change the cavity transmission~\cite{khudav2009}.

\begin{figure}[!ht]
  \centering
  \includegraphics[width=0.4\textwidth]{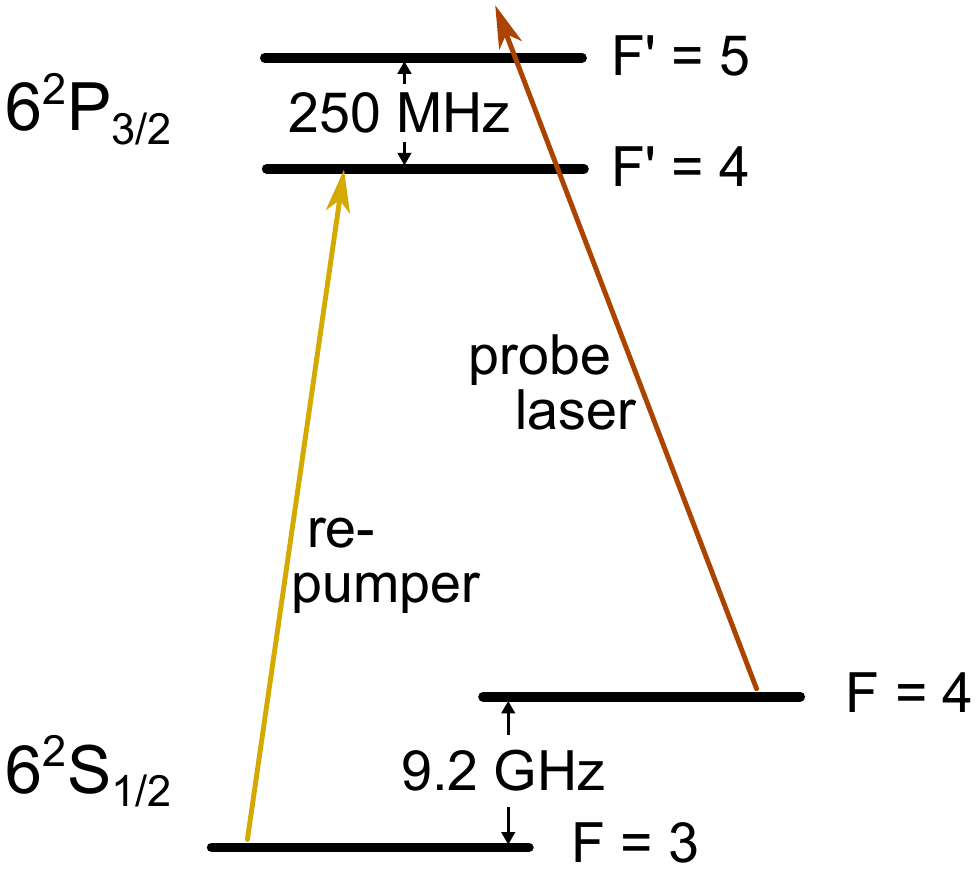}
  \caption{Simplified level scheme of the Caesium atom and the lasers used in the experiment.}
  \label{fig:scheme}
\end{figure}

\begin{figure}
\includegraphics[width=0.9\textwidth]{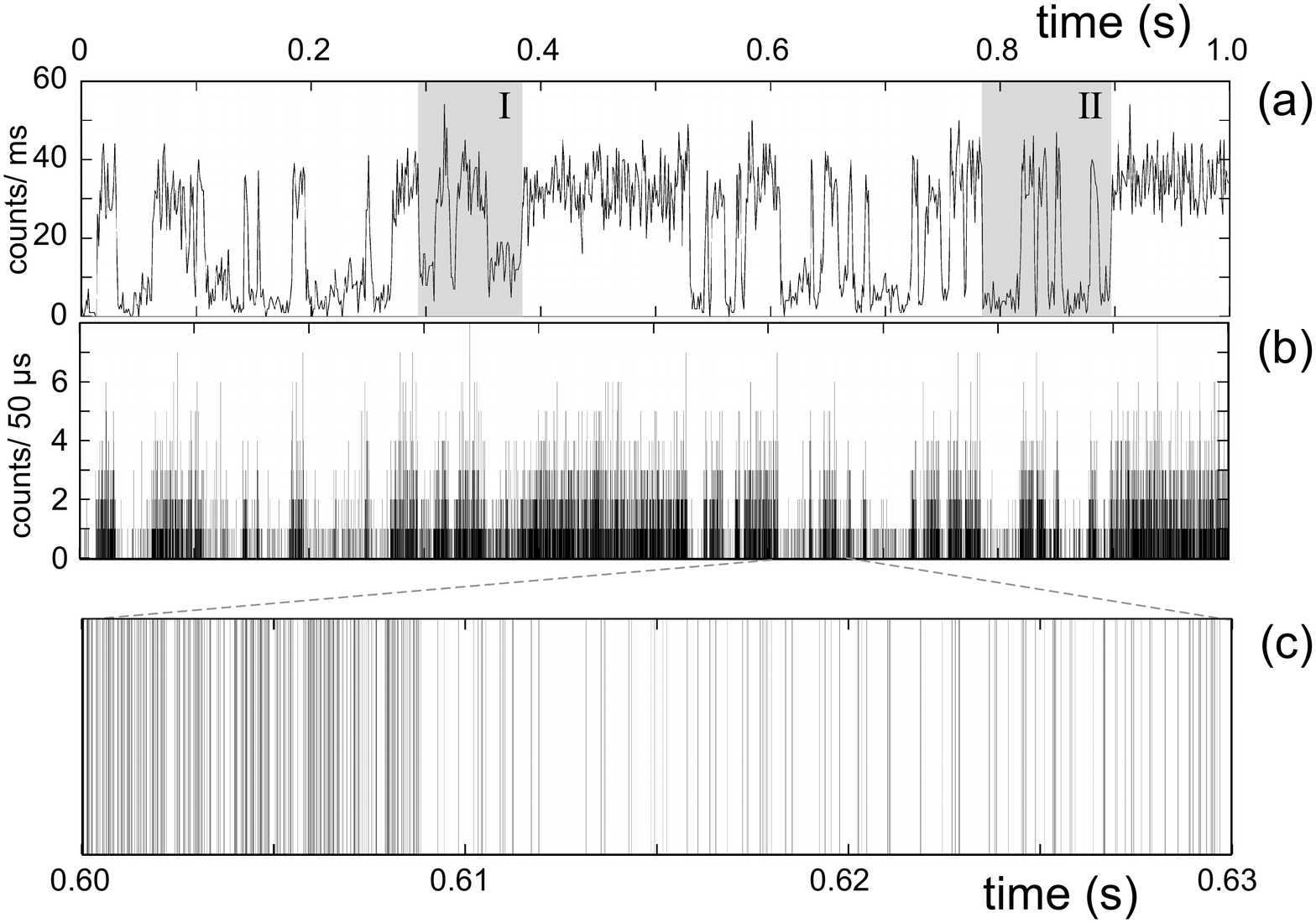}
 \caption{(a) Detected cavity transmission of photons binned in 1 ms time intervals. (b) The same data as in (a) binned in 50 $\mu$s time intervals. (c) Zoomed time region showing the individual detector clicks, experiencing a pronounced change in click rate as reflected also in (a) and (b).}
 \label{fig:data}
\end{figure}

The cavity resonance is stabilized 30\,MHz blue detuned to the $|F=4\rangle \rightarrow |F'=5\rangle$ transition of the D2 line at 852\,nm, see Fig.~\ref{fig:scheme}. A weak probe laser beam, resonant with the empty cavity, is coupled into the cavity and populates the cavity mode with 0.3 photons on average. In this blue-detuned weak probing regime, cavity cooling~\cite{ritsch1998} counteracts heating of the atom due to photon scattering and thus allows us to observe each atom over several seconds. The transmission of the probe light through the cavity is detected by a single-photon counting module with an overall detection efficiency of 4.4\%~\cite{reick2010}. The arrival time of each detected photon is recorded with 50\,ns time resolution, see Fig.~\ref{fig:data} and combined by software into time bins convenient for presentation of the data.

When we place an atom prepared in $|F=4\rangle$ into the cavity mode, the detected cavity transmission drops from about 30~counts/ms to 5...10~counts/ms, see Fig.~\ref{fig:data}(a), where the data is binned in 1 ms intervals. Within a few miliseconds, however, off-resonant scattering of the probe light fron the $|F'=4\rangle$ level transfers the atom to the $|F=3\rangle$ ground state with a rate of about 40\,s$^{-1}$. Since the optical transitions starting from $|F=3\rangle$ are detuned by the hyperfine splitting of $\Delta_\text{FHS}\approx 9.2\,\text{GHz}\gg (g,\kappa,\gamma)$ from the cavity resonance, the atom in this state does not noticeably influence the cavity resonance and the transmission raises to the high level of the empty cavity. The cavity transmission is thus a measure of the atom's hyperfine quantum state, and a quantum jump from $|F=4\rangle$ to $|F=3\rangle$ becomes visible as a sudden transition from low to high photon count rate.

A very weak laser beam tuned to the $|F=3\rangle \rightarrow |F'=4\rangle$ transition repumps the atom back to the $|F=4\rangle$ state with a rate comparable to the pumping rate due to the probe laser. In this way we obtain a sequence of quantum jumps between $|F=3\rangle$ and $|F=4\rangle$, and in the cavity transmission we observe the corresponding random telegraph signal~\cite{yuzhel2000}.

\section{Hidden Markov Modelling of cavity transmission experiment}

The physical process described in the previous chapter is a good candidate for a Hidden Markov Model analysis, i.e., as a stochastic process which produces a measurement signal $s_t$ whose statistics is governed by the current state $X_t$ of the (hidden) system which is itself governed by a Markov evolution process. In our case the signal $s_t$ is the number of photons recorded and the hidden system is the atom, whose state $X_t$ jumps randomly between the hyperfine states. The state and the signal are represented in discrete intervals of time, and for convenience we use integer values $t=1,2, ... $ to represent these bin time intervals of a finite duration $\Delta t$, which is short enough to accurately represent the atomic dynamics. The Markov property formally requires that the atomic state $X_t$ is conditionally independent of all variables except $X_{t-1}$ and that $s_t$ is conditionally independent of all variables except $X_t$. In Sec. V, we shall comment further on the Markovian Ansatz in connection with our modelling of the experiment.  Between measurements, the atomic system is governed by jumps between its internal states, as described by the \emph{transition} probabilities, $P(X_{t+1}|X_t)$. The signal, on the other hand, is governed by a conditional probability distribution $P(s_t|X_t)$. In this section we show how recursive application of Bayes' rule, $P(A|B) = P(B|A)P(A)/P(B)$,  permits calculation of the probability for the atomic state, conditioned on the measured data. In Sec. IV, we show how the calculation can be applied in an iterative manner to re-estimate also the jump rates and the signal conditional probabilities, which are assumed to be known input to the Bayesian analysis.

The joint probability for the stochastic variables $X$ and $s$ to have attained particular values at all instances of time, $t=1, .. , N$,  can be written as a product of conditional probabilities and a prior atomic state probability,
\begin{align}
 P(X_1, \ldots X_N, s_1, \ldots s_N) = \prod_{j=1}^{N} P(s_j | X_j) \prod_{j=1}^{N-1} P(X_{j+1} | X_j) P(X_1). \label{eq:full_joint}
\end{align}

We recall that the optical transmission signal $s_t$ is given by the experiment, while the atomic state $X_t$ is not directly observed, and the goal of the present work is to determine the probability distribution $P(X_t)$ conditioned on the experimental results as sharply as possible.

\subsection{Forward Bayesian state estimate}

Let us first present the conventional situation, where one uses the signal data to estimate the current state of the system, by a recursive Bayesian update. We shall refer to this as a "forward estimate", because the data acquired until  a given instant $t$ is used to estimate the state at time $t$.

While the current state probabilities depend on the entire signal prior to the current time, they need not be formally computed as a function of these many variables, as the distribution can be updated at every instant of time taking into account only the prior probability and the most recently measured result.
We assume an initial prior $P(X_1)$ for the atomic state and we note that given $P(X_t | s_1, \ldots s_t)$ the probability distribution at the subsequent time is given by the transition probabilities,
\begin{align}
 P(X_{t+1}| s_1, \ldots s_t) = \sum_{X_t} P(X_{t+1}|X_t) P(X_t|s_1, \ldots s_t)
\end{align}
This provides the prior for the Bayesian estimate of the state at time $t+1$, and  the probability conditioned on the measurement result $s_{t+1}$ becomes
\begin{align}
 P(X_{t+1}| s_1, \ldots s_t, s_{t+1}) &\propto P(s_{t+1}|X_{t+1}) P(X_{t+1}|s_1, \ldots s_t) \nonumber\\
  &= \sum_{X_t} P(s_{t+1}|X_{t+1}) P(X_{t+1}|X_t) P(X_t|s_1, \ldots s_t) \label{eq:recursive_bayes}
\end{align}
where the constant of proportionality can be found by normalization. The combined effect of transition probabilities and Bayesian conditional probabilities is repeated at every time step, and such a procedure was succesfully applied in \cite{reick2010} to the case of both one and two atoms inside an optical cavity, and its achievements were compared favorably to the simpler state estimation based on the noisy instantaneous transmission signal in different time bins.

\subsection{Full Bayesian state estimate}

Any information that we acquire, which contains further information about the atomic state will lead to an improved estimate and a conditional update of $P(X_t)$. In this section we will focus on the information about $X_t$ available in the measurement signal $s_{t+1},s_{t+2}, ... s_N$ \textit{after} time $t$. This information must be dealt with via appropriate definition and calculation of conditional probabilities and transition probabilities.

Let us first, however, discuss the meaning of assigning probabilities and making predictions for past events. Clearly, we can not go back in time and verify the prediction, and if we already measured the atomic state at time $t$, it would already be known to us and no further refinement of $P(X_t)$ would be relevant. Note, however, that in the case of classical probabilities, we assign the probabilities to states that could have been measured and the result could have been copied and stored, so that they could be compared with our later predictions. A meaningful test of "past predictions" could then  involve the comparison of our prediction with the actual outcome of a measurement which had a result that was not revealed at the time of the measurement.

Time dependent state estimation based on a full detection record constitutes, indeed, an already well established component, called smoothing, in the theory of Hidden Markov Models.
For completeness we provide a brief summary of this method, following the presentation in \cite{NumericalRecipes3rd}. First, it is convenient to introduce two time dependent quantities, a joint and a conditional probability distribution,
\begin{align} \label{alphabeta}
 \alpha_t(i) &= P(X_t = i,s_1, \ldots s_t) \\
 \beta_t(i) &= P(s_{t+1}, \ldots s_N | X_t = i).
\end{align}

By normalization of $\alpha_t(i)$, we recognize the forward state estimate,
\begin{align} \label{eq:alpha_forward_estimate}
 P(X_t = i| s_1, \ldots s_t) &= \frac{P(X_t = i, s_1, \ldots s_t)}{P(s_1, \ldots s_t)}
    = \frac{\alpha_t(i)}{\sum_k \alpha_t(k)},
 \end{align}
while $\alpha_t(i)$ and $\beta_t(i)$ together serve to provide the state probabilities, conditioned on the full measurement record
 \begin{align} \label{eq:alpha_beta_estimate}
 P(X_t = i| s_1, \ldots s_N) &= \frac{P(X_t = i, s_1,\ldots s_N)}{P(s_1, \ldots s_N)} \nonumber \\
    &= \frac{ P(s_{t+1}, \ldots s_N|X_t = i) P(s_1, \ldots s_t, X_t = i) }{P(s_1, \ldots s_N)}
    = \frac{\alpha_t(i) \beta_t(i)}{\sum_k \alpha_t(k) \beta_t(k)}.
\end{align}
Here, we have used that $\alpha_t(i) \beta_t(i) = P(X_t = i, s_1, \ldots s_N)$ and that the probability for observing $s_1, \ldots s_N$ \textit{and} $X_t = i$ is the probability for observing $s_1, \ldots s_t$ and $X_t = i$ multiplied by the probability for observing $s_{t+1}, \ldots s_N$ \emph{given} $X_t = i$.

It follows from the Markov property that
 \begin{align}
  \alpha_{t+1}(i) = \sum_{j} P(s_{t+1}|X_{t+1} = i ) P(X_{t+1} = i |X_t = j) \alpha_t(j). \label{eq:alpha_update}
\end{align}

To obtain a similar update rule for $\beta_t(i)$, we consider
\begin{align}
 P(s_{t+1}, \ldots s_N | X_{t+1}) = P(s_{t+1}|X_{t+1}) P(s_{t+2}, \ldots s_N | X_{t+1}).
\end{align}
From which we obtain
\begin{align}
 P(s_{t+1}, \ldots s_N | X_t=i) &= \sum_j P(s_{t+1}, \ldots s_N | X_{t+1}=j) P( X_{t+1}=j | X_t=i )\nonumber  \\
  & = \sum_j P(s_{t+1}|X_{t+1}=j) P(X_{t+1}=j | X_t=i) P(s_{t+2}, \ldots s_N | X_{t+1}=j),
\end{align}
where the first equality follows from the general formula $P(x|z) = \sum_y P(x|yz)P(y|z)$, with $x = s_{t+1}, \ldots s_N$, $y = X_{t+1}=j$ and $z = X_t=i$ \emph{and} the fact that in our case $P(x|yz) = p(x|y)$ due to the Markov property.

We have thus obtained the desired recursive equation for $\beta_t(i)$:
\begin{equation}
\beta_t(i) = \sum_j P(s_{t+1}| X_{t+1}=j) P( X_{t+1}=j | X_t=i ) \beta_{t+1}(j),\label{eq:beta_update}
\end{equation}
and even though $\beta_t$ represents a seemingly complicated conditonal probability for the entire signal sequence, the equation is readily solved recursively backwards from the final instant of measurements where $\beta_N(i)=1$.

Finally, combining the solution for $\alpha$ and $\beta$, we have the state estimate based on the full detection record (\ref{eq:alpha_beta_estimate}).

\section{Re-estimating parameters}

It will often be the case that some physical parameters are not precisely known, and that they will have to be determined by the same or a similar experiment as the one revealing the quantum state itself. In the Bayesian formalism, it is possible to treat not only the quantum state but also the value of a small set of variables probabilistically. Such parameter estimation, which is possible if the parameter space is not too large, was demonstrated in \cite{Brakhane-12} in a simultaneous determination of the atomic state and the transition rates between the atomic states from the photon transmission data set. For a larger number of unknown parameters, the multidimensional Bayesian filter becomes numerically intractable. One may then have recourse to the more direct approaches discussed in this section.

\subsection{Re-estimation of transition probabilities}

Consider first the case where we wish to determine the atomic transition probabilities $P(X_{t+1}=j|X_t=i)$ for the transition from state i to state j. We assume that this probability is time-independent, and we therefore write it as $P(X_{+1}=j|X=i)$ in the following. The transitions are revealed through the time-dependent probability that we occupy state $i$ and $j$ at subsequent time steps, which are in turn, inferred from the measured data,
\begin{align}
 \gamma_t(i, j) &\equiv P(X_t = i, X_{t+1} = j| s_1, \ldots s_N) = \frac{P(X_t = i, X_{t+1} = j, s_1, \ldots s_N)}{P(s_1, \ldots s_N)}
\end{align}
Now,
\begin{align}
 P(X_t = i, X_{t+1} = j, s_1, \ldots s_N)\\ &= P(X_t = i, s_1, \ldots s_t) P(X_{t+1}=j|X_t = i) P(s_{t+1}|X_{t+1}=j) P(s_{t+2}, \ldots s_N| X_{t+1} = j) \\
  &= \alpha_t(i) P(X_{t+1}=j|X_t = i) P(s_{t+1}|X_{t+1}=j) \beta_{t+1}(j)
\end{align}
Since  $P(s_1, \ldots s_N) = \sum_k \alpha_t(k) \beta_t(k)$ we may also  write
\begin{align}
 \gamma_t(i, j)  = \frac{\alpha_t(i) \beta_{t+1}(j)}{\sum_k \alpha_t(k)\beta_t(k)}\cdot P(X_{t+1}=j|X_t = i) P(s_{t+1}|X_{t+1}=j)
\end{align}

This enables the re-estimation of the transition probabilities $P_{est}(X_{+1}|X)$
\begin{align}
 P_{est}(X_{+1}=j|X = i) = \frac{\sum_t \gamma_t(i, j)}{\sum_t (\sum_j \gamma_t(i, j)) } = \frac{\sum_t \gamma_t(i, j)}{\sum_t P(X_t = i|s_1, \ldots s_N) }, \label{reest_trans}
\end{align}
where the second equality follows from
\begin{align}
 \sum_j \gamma_t(i, j) = \frac{\alpha_t(i) \beta_t(i)}{\sum_k \alpha_t(k)\beta_t(k)} = P(X_t = i|s_1, \ldots s_N).
\end{align}
We recognize that Eq.(\ref{reest_trans}) in a probabilistic manner counts the number of times a transition occurs between $i$ and $j$ divided by the accumulated occupation of the  $i$-state.

Note, that for the calculation of this quantity, the forward-backward estimated state provides a more accurate estimate of the parameters: Since the estimated state is more accurate, the counting of the number of transitions $i \to j$ is more accurate, which leads to the more accurate estimate of the transition probabilities.

\subsection{Re-estimation of signal probabilities}

As shown in \cite{reick2010}, the actual photon counting distribution for a given known atomic state may be broader than a Poisson distribution, possibly associated with the physical motion of the atom between locations that experience different coupling strengths to the cavity field mode. Assuming, for simplicity, that the counting distribution is governed by the Markov property and depends conditionally only on the atomic state, a simple procedure determines $P(s|X)$ by counting how often the signal outcome $s$ occurs when the atom is in state $X$. Since the occupation of state $X=i$ is given probabilistically, this translates into the expression,
\begin{align}
P(s| X = i ) = \frac{ \sum_t P(X_t = i|s_1, \ldots s_N) \delta(s - s_t)}{\sum_t P(X_t = i|s_1, \ldots s_N) }. \label{eq:EmissionReEstimation}
\end{align}

The above procedure is the so-called Baum-Welch estimation which can be shown \cite{NumericalRecipes3rd} to be equivalent to the so-called estimation-maximization (EM) algorithm, and will always converge to a (local) optimal estimate. There is, however, no guarantee that the procedure will find the globally optimal model. As in the determination of the atomic transition rates, the better the knowledge of the state occupation in state $X_t=i$, the better is the estimation of the signal probability distribution. As we shall show with our numerical example, the state estimation based on the full data record is far superior over the only forward Bayesian method, and therefore the parameter estimation is also expected to be much more precise.

Note that the equations in this and the previous section have to be iterated, since the conditional signal probabilities determine the state estimate, which in turn leads to an updated estimate of the signal probabilities. Further details of this iterative scheme and on the identification of a suitable Hidden Markov Model will be described in the next section.

\section{Hidden Markov Model analysis of experiments}

In this section we apply the procedures reviewed in the previous sections to the data obtained in the experiments described in Sec. II. While the atom has two ground hyperfine ground states, and would thus seem amenable to an HMM description with two (hidden) states, it was observed in \cite{reick2010} that the atom in the $|F=4\rangle$ state showed super-Poissonian counting statistics, Var$(n) > \langle n\rangle)$. In \cite{reick2010} this was ascribed to the influence of other degrees of freedom of the atom, e.g., its spatial position, which would affect its coupling to the cavity field mode. If such degrees of freedom in the system are evolving on a significantly longer time scale, it would violate the basic Markov assumption of HMM to treat the purported $|F=4\rangle$ occupation by a single hidden state. Such an effect can be clearly seen in Fig. 3(a) when comparing the transmission level around 0.3 to 0.4 s and around 0.8 to 0.9 s. The HMM theory outlined in Secs. III and IV is, however, ideally suited to problems with an unknown number of hidden states with unknown dynamics. It is the basic idea of the method that it allows "black box" modelling of an observed signal by states, transition rates and signal properties. We will thus allow more than two states in our modelling of the system, and use the state and parameter estimation to assess their physical meaning afterwards.

Fig.~\ref{fig:data} is, indeed, suggestive of a dynamical evolution of the system between different levels of low transmission, and we shall therefore attempt a model of the system with three states. We do not restrict the properties of these states, but we expect that the algorithm will use the available parameter space to represent one non-interacting state $|F=3\rangle$ and two different states belonging to $|F=4\rangle$ with slightly different transmission characteristics.

\begin{figure}
\includegraphics[width=0.9\textwidth]{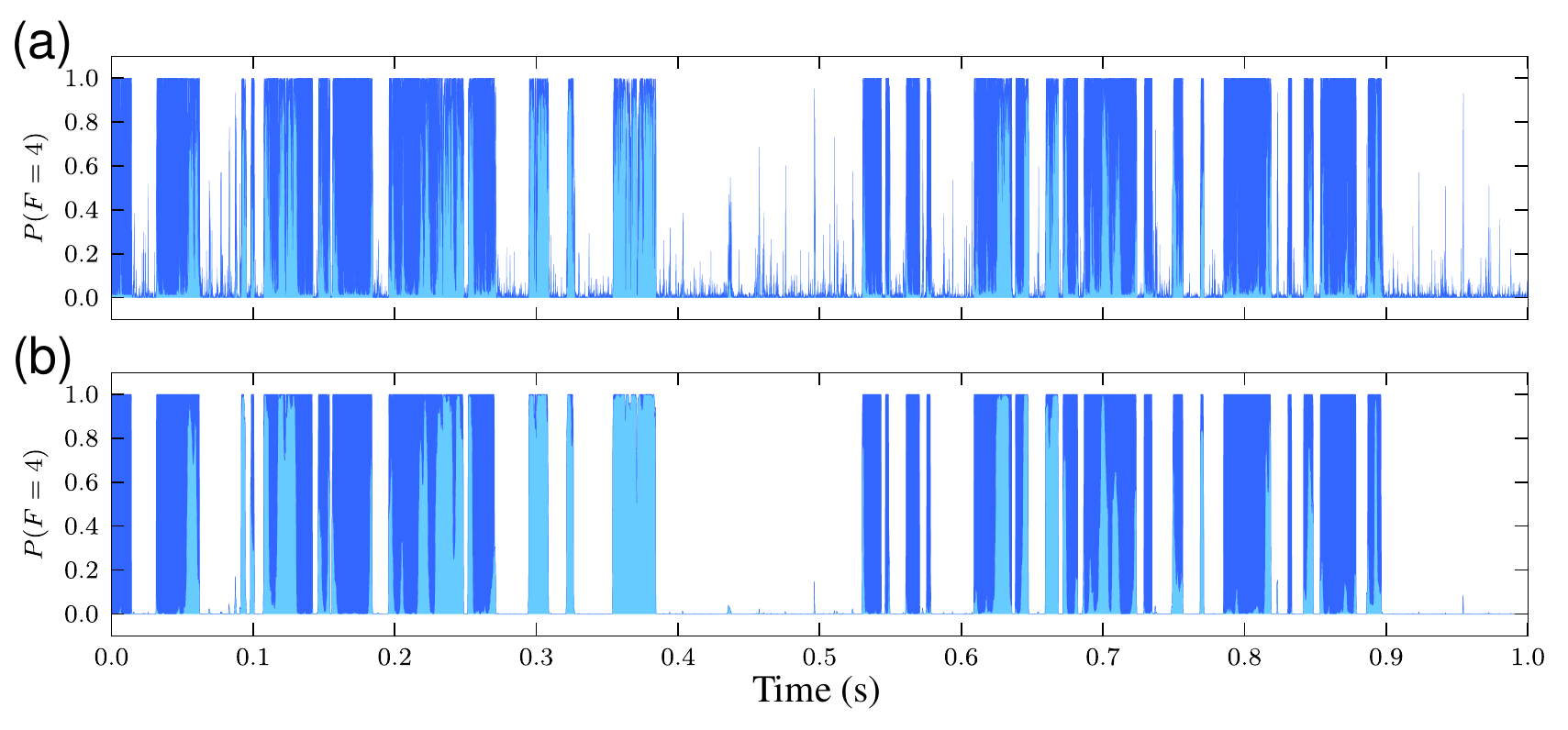}
 \caption{(a) The time dependent probability for the atom to be in $|F=4\rangle$, inferred from the forward Bayesian analysis of the measured data. The two states with slightly different transmission characteristics hidden in $|F=4\rangle$ are indicated with dark and light blue color. (b) The atomic state populations  conditioned on the forward-backward Bayesian state estimation. In both figures, the state is conditioned on the data shown in Fig.~\ref{fig:data}} \label{fig:HMMStateEstimate}
\end{figure}

As described above, we first use the Bayesian state update formulas with a more or less random initial Ansatz for the rate and signal parameters, and we then subsequently iterate the re-estimation and state estimation process until they have converged. We have tested that this procedure converges to the same values with different initial parameter choices, and the results we show in this section are the ones obtained after consistently identifying the model parameters by this iterative procedure.

The forward Bayesian estimation of the atomic state applies the recursive update formula (\ref{eq:recursive_bayes}) for every 50 $\mu$s time bin of the data acquisition, and this determines the time dependent state of the system, shown in the upper panel in Fig.~\ref{fig:HMMStateEstimate}. In reality, the HMM method identifies the time dependent population of three different states, but it does not identify which are these hidden atomic state. Our estimation of the signal rates associated with the states, however, clearly yield the expected outcome: two hidden states comply with two slightly different low transmission levels and are thus identified with the atom in the $|F=4\rangle$ hyperfine state, and one state complies with the high transmission of the atomic $|F=3\rangle$  state. In Fig. \ref{fig:HMMStateEstimate}, we plot the two $|F=4\rangle$ populations by the light and dark blue shadings, and we observe that most of the time they add to values near zero and unity while, their individual populations fluctuate more. This confirms our intuition that the transmission signal clearly distinguishes high from low transmission, while statistical fluctuations prevent a clear distinction between the two states which are assigned low but quite similar transmission.

The measured signal leads to a number of narrow spikes both in the individual and in the total $|F=4\rangle$ populations. Most of the narrow spikes clearly correspond to statistically improbable short intervals spent by the atom in one and the other hyperfine state, and they result from the inability of the forward Bayesian update formula to distinguish a short time statistical fluctuation in the transmission signal from an actual change due to the transfer of the atom into a different state. The forward-backward estimate, however, conditions the atomic state on the full measurement record according to Eqs.(\ref{eq:alpha_beta_estimate},\ref{eq:alpha_update},\ref{eq:beta_update}), and the results are shown in Fig. 4(b). Here, indeed, the atomic state populations in the $|F=3\rangle$ and the $|F=4\rangle$ states are significantly closer to zero and unity at all times, and most of the spikes in the upper panel of the figure have disappeared. The reason for this is the use of the full detection record, in which, e.g., a brief dip in the signal is recognized as a statistical fluctuation and thus distinguished from an actual change from high to low transmission. The figure is a clear proof of what we set out to demonstrate: the forward-backward Bayesian estimate provides much more decisive predictions for the atomic state than the usual conditioned dynamics. The forward-backward analysis also distinguishes sharper between the two candidate $|F=4\rangle$ states: the data relevant to the state assigment at any time is approximately doubled when both the past and the future transmission is considered.

Naturally, Fig.~\ref{fig:HMMStateEstimate} only compares two sets of predictions with each other, and to test theory and confirm its predictions, measurements should be carried out. One simple test of the theory could consist in hiding the data acquired in a single time bin and compare it later with the theoretical predictions, which are governed by $P(s_t)=\sum_i P(s_t|X_t=i)P(X_t=i)$. The state probabilities $P(X_t=i)$ are assigned different values by the forward and the forward-backward recursive expressions (\ref{eq:recursive_bayes},\ref{eq:alpha_beta_estimate}), and our improved predictions about the past state of the atom implies an improved ability to guess the photon count in the past time bin.

\begin{figure}
  \includegraphics[width=0.9\textwidth]{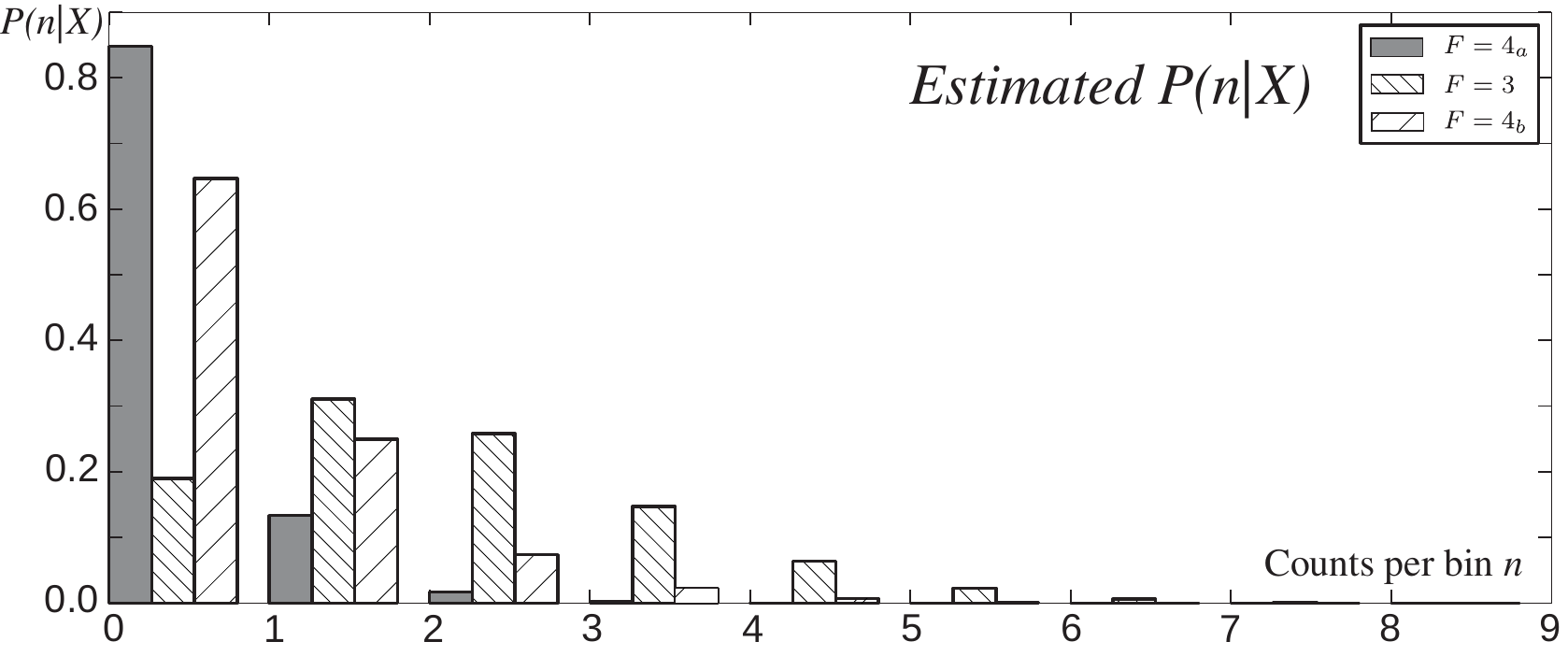}
 \caption{
 (Color online) State dependent photon count distribution for a time of 50 $\mu$s obtained after 200 iterations of the re-estimation algorithm described in the text. The photon count distribution with high mean value (middle, down hatched columns) is associated with the $|F=3\rangle$ state, while the HMM algorithm yields two states with low transmission that we ascribe to the $|F=4\rangle$ hyperfine state (left and right, shaded and up hatched columns).} \label{fig:HMMParameters}
\end{figure}

As stated in the beginning of this section, we have iteratively estimated both the atomic state and the transition and transmission probabilities. For the photon transmission, we  have applied this recursive scheme to the data from the experiments, assuming  that the number $n$ of detector clicks in every 50 $\mu$s time bin has an unknown probability distribution, $P(n|X)$ that we update according to Eq.~(\ref{eq:EmissionReEstimation}) until the conditioned state dynamics and the parameter re-estimation converged. After about 200 iterations, all rates and signal probabilities were converged to within a tolerance of \num{1e-9}, and we arrived at the distributions shown in Fig.~\ref{fig:HMMParameters}.

\noindent

The histograms shown in Fig.~\ref{fig:HMMParameters}, represent the actual counts in 50 $\mu$s intervals, reported in Fig.~\ref{fig:data}(b), sorted according to the state assignment in Fig.~\ref{fig:HMMStateEstimate}(b). The assignment of two different signal probabilities to the $|F=4\rangle$ state is compatible with the super-Poissonian character in the low count signals, and we observe that the two distributions are sufficiently different to be distinguished if the atom resides in one or the other states for long enough. This indeed happens several times during the measurement sequence where the forward-backward state assignment favors one or the other of the $|F=4\rangle$ states with near unity probability.

Our analysis also determines the rates of transitions between the three different states in the model. Since the Baum-Welsch expression for the transition rates makes explicit use of the time dependent conditional probability assignments, we see in Fig.~\ref{fig:HMMStateEstimate} that the forward Bayes analysis with its many sharp jumps will strongly overestimate the atomic transition rates, while the forward-backward analysis yields much fewer transitions. Indeed, the rates for transitions from the $|F=3\rangle$ to the $|F=4\rangle$ states (and back) are estimated to $\sim$ 60 ($\sim$ 80) transitions per second, in better agreement with estimates based on the laser-atom parameters made in Sec. II.

\section{Discussion}

We have presented a Hidden Markov Model for cavity transmission experiments with an atom jumping between different states. We showed that the standard HMM formalism, taking only the experimental data as input was able to identify states with the expected characteristics and a dynamical jump process between states in agreement with our physical understanding of the dynamics. This lends support to the validity of the state assignment, shown in Fig.4, and it brings promises for use of the same method to analyze a wider range of conditional dynamics experiments.

We wish, in particular to highlight the use of the forward-backward estimate which, by incorporating the whole data set rather than only data taken prior to the time at which the state is determined, provided a considerable improvement of the state analysis. Due to its significantly sharper predictions, the forward-backward analysis yields also better estimates of the physical process parameters, rates and signal probabilities. It is already a well established procedure in classical HMM theory, and it should become a standard component in the analysis of quantum dynamics.

In this paper, the atomic system underwent only incoherent processes, and the conditional dynamics was described by classical probability theory. We have recently shown \cite{Gammelmark2013} that a full forward-backward theory can also be established for coherent quantum dynamics of a system subject to repeated or continous measurement. In that theory, predictions for the outcome of quantum measurements are given by two matrices, which become diagonal with elements equal to the $\alpha_t(i)$ and $\beta_t(i)$ in Eq.(7) in the case of incoherent processes.

Let us, finally, comment on the choice of three states for the HMM theory. This choice was motivated by the observation that only two states would not be compatible with the assumption of Markovian dynamics (the atom in the $|F=4\rangle$ appeared to have a "memory" leading to a slowly varying transmission at the low transmission level). With three states, the model identifies two underlying states, possibly associated with the atomic motion inside the cavity. We have tested the Ansatz by further augmenting the model and allowing also four hidden states. That calculation led to a further splitting of the state with $|F=3\rangle$ transmission characteristics. The pair of states obtained this way, however, show very similar transmission probabilities, and the 4-state model does not appreciably change the assignment of (total) time-dependent probabilities of occupation of the $|F=3\rangle$ and the $|F=4\rangle$ hyperfine states compared with the 3-state model. We thus conclude that the hyperfine state assignment shown in Fig.4 is robust to refinements and changes in the modelling.

The authors wish to acknowledge financial support jointly from the CCQED training network of the European Commission. Aarhus also wishes to
thank the Villum Foundation and the Aarhus University Research Foundation. Bonn acknowledges support by the Bundesministerium für Forschung und Technologie
(BMFT, Verbund QuOReP), and the Integrated Project SIQS funded by the European Commission.

\end{document}